\documentclass[%
 reprint,
 superscriptaddress,
 amsmath,amssymb,
 prx,longbibliography
]{revtex4-1}
\usepackage{graphicx}
\usepackage{dcolumn}
\usepackage{bm}
\usepackage{float}
\usepackage{siunitx}
\usepackage{color}
\usepackage{amsmath}
\usepackage{mathtools}
\usepackage[colorlinks]{hyperref}
\hypersetup{%
    plainpages=true,
    breaklinks=true,
    hypertexnames=false,
    pageanchor=true,
    colorlinks=true,
    linkcolor={black},
    citecolor={blue},
    urlcolor={blue},
    anchorcolor={black}
}

\usepackage[utf8]{inputenc}
\usepackage{dsfont}
\usepackage{cleveref}
\usepackage{xcolor}
\usepackage{pgf}
\usepackage{placeins}

\usepackage{newfloat}
\begin{document}

\preprint{APS/123-QED}

\title{On-Demand Directional \textcolor{black}{Microwave} Photon Emission \\Using Waveguide Quantum Electrodynamics}

\def\RLEaffil{Research Laboratory of Electronics, Massachusetts Institute of Technology, Cambridge, MA 02139, USA}
\def\LLaffil{MIT Lincoln Laboratory, Lexington, MA 02421, USA}
\def\Physaffil{Department of Physics, Massachusetts Institute of Technology, Cambridge, MA 02139, USA}
\def\EECSaffil{Department of Electrical Engineering and Computer Science, Massachusetts Institute of Technology, Cambridge, MA 02139, USA}

\author{Bharath Kannan,$^{1,\,2,\,*,\,\dagger,\,\ddagger}$ Aziza Almanakly}
\thanks{These authors contributed equally.\\$^{\dagger}$ bkannan@mit.edu \\ $^{\ddagger}$ Present address: Atlantic Quantum, Cambridge, MA 02139, USA}
\affiliation{\RLEaffil}
\affiliation{\EECSaffil}

\author{Youngkyu Sung,$^{1,\,2,\ \ddagger}$ Agustin Di Paolo}
\affiliation{\RLEaffil}


\author{David A. Rower}
\affiliation{\RLEaffil}
\affiliation{\Physaffil}

\author{Jochen Braum\"uller}
\affiliation{\RLEaffil}

\author{Alexander Melville}
\author{Bethany M. Niedzielski}
\affiliation{\LLaffil}

\author{Amir Karamlou}
\affiliation{\RLEaffil}
\affiliation{\EECSaffil}

\author{Kyle Serniak}
\affiliation{\LLaffil}

\author{Antti Veps\"al\"ainen}
\affiliation{\RLEaffil}

\author{Mollie E. Schwartz}
\author{Jonilyn L. Yoder}
\affiliation{\LLaffil}

\author{Roni Winik}
\author{Joel I-Jan Wang}
\affiliation{\RLEaffil}

\author{Terry P. Orlando}
\affiliation{\RLEaffil}
\affiliation{\EECSaffil}

\author{Simon Gustavsson,$^{1,\,\ddagger}$ Jeffrey A. Grover}
\affiliation{\RLEaffil}
 
\author{William D. Oliver}
\affiliation{\RLEaffil}
\affiliation{\EECSaffil}
\affiliation{\Physaffil}
\affiliation{\LLaffil}


\begin{abstract}
Routing quantum information between non-local computational nodes is a foundation for extensible networks of quantum processors.
Quantum information transfer between arbitrary nodes is generally mediated either by photons that propagate between them, or by resonantly coupling nearby nodes.
The utility is determined by the type of emitter, propagation channel, and receiver.
Conventional approaches involving propagating microwave photons have limited fidelity due to photon loss and are often unidirectional, whereas architectures that use direct resonant coupling are bidirectional in principle, but can generally accommodate only a few local nodes.
Here we demonstrate high-fidelity, on-demand, directional, microwave photon emission. We do this using an artificial molecule comprising two superconducting qubits strongly coupled to a bidirectional waveguide, effectively creating a chiral microwave waveguide.
Quantum interference between the photon emission pathways from the molecule generates single photons that selectively propagate in a chosen direction.
This circuit will also be capable of photon absorption, making it suitable for building interconnects within extensible quantum networks.
\end{abstract}
\maketitle

\section{Introduction}
Most realistic architectures of large-scale quantum processors employ quantum networks: the high-fidelity communication of quantum information between distinct non-local processing nodes~\cite{Kimble2008}.
Quantum networking enables modular and extensible quantum computation by mediating distributed entanglement between computational nodes~\cite{Cirac1997,Cirac1999}.
There are several approaches to realizing quantum networks, including the routing of optical photons between trapped-ion modules~\cite{Monroe2014}, coupling emitters to photonic waveguides~\cite{Sollner2015, Coles2016} or optical nanofibers~\cite{Petersen2014, Mitsch2014, Lodahl2017, Solano2017}, shuttling ions~\cite{wan2019,Pino2021} or neutral atoms~\cite{bluvstein2021quantum} between qubit arrays, or cavity-assisted pairwise coupling between natural or solid-state artificial atoms~\cite{Zhong2019,Leung2019,Chang2020,Zhong2021,Burkhart2021,ramette2021}.
Enabling non-local quantum communication is particularly relevant for qubits which are natively limited to nearest-neighbour coupling, such as 2D arrays of surface-trapped ions, semiconducting qubits, and superconducting qubits. 

Experimental realizations of communication between superconducting qubits have typically relied on coherent coupling via resonators~\cite{Zhong2019,Leung2019,Chang2020,Zhong2021,Burkhart2021} or itinerant photons that propagate in unidirectional waveguides\textcolor{black}{~\cite{Kurpiers2018, Axline2018, Campagne-Ibarcq2018, Kurpiers2019,Magnard2020}}.
While the former approach has achieved the highest fidelities to date, it is not easily extensible.
For example, the free spectral range of the coupling resonator constrains the maximal distance between the nodes.
Alternatively, itinerant photons that propagate along waveguides do not have this limitation.
However, the fidelity of this approach has been limited as lossy non-reciprocal components, such as circulators, are required to prevent undesirable standing waves between nodes and render waveguides---that are naturally bidirectional---unidirectional.
Instead, an architecture that uses conventional bidirectional waveguides, in conjunction with the ability to generate photons which propagate in a chosen direction, would enable both high-fidelity and high-connectivity communication within a quantum network.

Recent theoretical work has shown that superconducting circuits in a waveguide quantum electrodynamics (QED) architecture are capable of realizing such a network\textcolor{black}{~\cite{Gheeraert2020, Guimond2020, Solano2021}}.
In waveguide QED, atoms are directly coupled to the continuum of propagating photonic modes in a waveguide~\cite{Lalumiere2013}.
Realizing the strong coupling regime of waveguide QED has enabled a wide range of phenomena to be experimentally observed, such as 
resonance fluorescence~\cite{Astafiev2010,Hoi2011, Hoi2013,Hoi2015}, 
Dicke super- and sub-radiance~\cite{Dicke1954,VanLoo2013,Mirhosseini2019}, 
and giant artificial atoms~\cite{FriskKockum2014,Kockum2018,Kockum2021review,Vadiraj2021,kannan2020}.

Importantly, the achievable strong coupling between superconducting qubits and itinerant photons enables the qubits to be used as high-quality quantum emitters~\cite{Abdumalikov2011,Hoi2012,Forn-Diaz2017,Gonzalez-Tudela2015,Pfaff2017,Gasparinetti2017,besse2020realizing,kannanN00N2020}.
\textcolor{black}{Spatial-mode matching remains a challenge with optical emitters, such as neutral atoms near optical nanofibers~\cite{Corzo2019, Solano2019}.
One can instead engineer the bandgap of a photonic crystal waveguide to achieve coupling efficiencies of up to $50\%$ with neutral atoms \cite{Goban2015} and $99\%$ with optical quantum dots \cite{Scarpelli2019}.
With superconducting circuits, however, qubit-waveguide coupling efficiencies greater than $99 \%$ are readily accessible without the need for slow-light waveguides or field enhancement from cavities~\cite{Mirhosseini2019, kannanN00N2020}.
}
\begin{figure*}[t!]
    \centering
    \includegraphics[width=\textwidth]{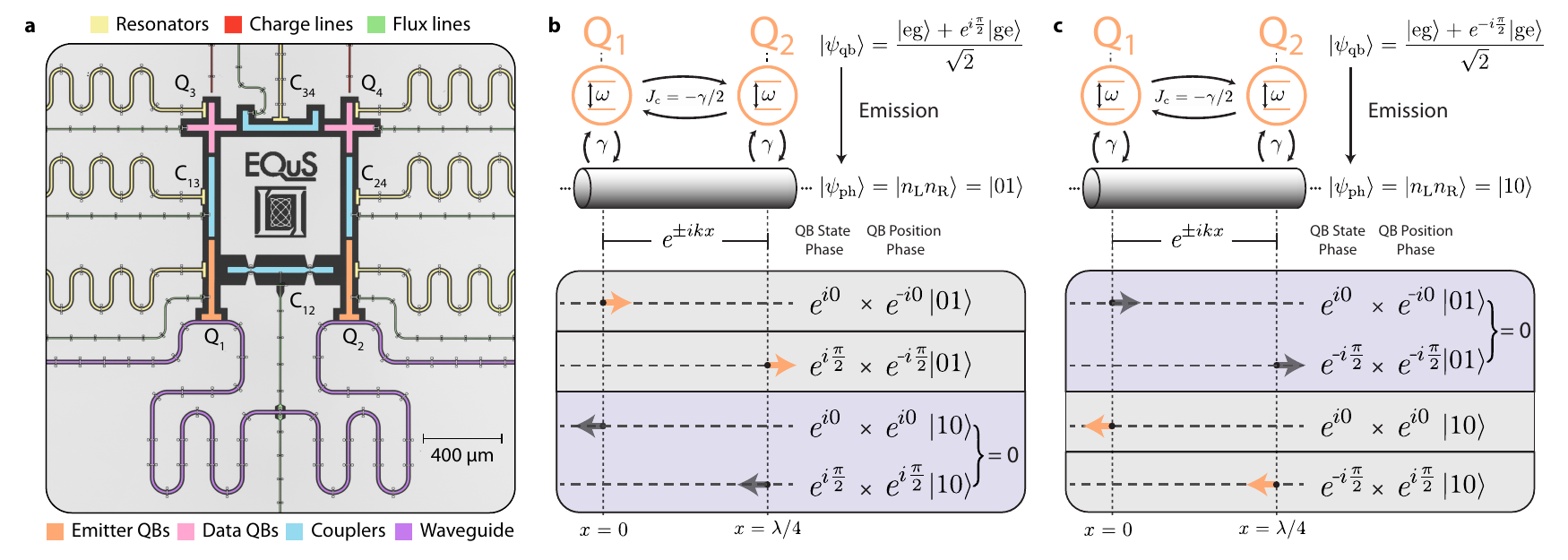}
    
    \caption{
    \textbf{Directional emission in a waveguide QED architecture.}
    \textbf{a)} A false-colored optical micrograph of the device.
    The state of the data qubits (pink) is transferred into the emitter qubits (orange) via an exchange interaction mediated by tunable couplers (blue).
    The emitter qubits continuously emit any population into the waveguide (purple).
    \textbf{b)} Schematic diagram of the two resonant emitter qubits $\textrm{Q}_1$ and $\textrm{Q}_2$ coupled to a common waveguide with equal strength $\gamma$ and separated by a distance $\lambda/4$.
    The phase delay for photons in the waveguide is given by $e^{\pm ikx}$, where $k = 2\pi/\lambda$ is the photon wavevector and $\lambda$ is the photon wavelength.
    The sign of this phase delay is determined by the propagation direction of the photon ($+$ for leftward, and $-$ for rightward).
    An external coupler-mediated exchange interaction of strength $J_\textrm{c}=-\gamma/2$ is applied in order to fully cancel the waveguide mediated interaction between the qubits.
    The four possible coherent pathways for a photon to be emitted by the qubits into the left/right travelling modes of the waveguide are shown below.
    Each pathway obtains a phase from the initial state $|\psi_\textrm{qb}\rangle$ and position $x$ of the qubit that is emitting a photon.
    When the qubits are initialized into $|\psi_\textrm{qb}\rangle = (|eg\rangle + e^{i\pi/2}|ge\rangle)/\sqrt{2}$, the emitted photon only propagates towards the right due to destructive interference between the left-propagating pathways.
    \textbf{c)} The same setup as (b), but with the initial qubit state $|\psi_\textrm{qb}\rangle = (|eg\rangle + e^{-i\pi/2}|ge\rangle)/\sqrt{2}$.
    In this case, the right-propagating pathways destructively interfere, and the emitted photon only propagates towards the left.}
    \label{fig:fig1}
\end{figure*} 

\textcolor{black}{In recent years, directional emission into a waveguide has 
become a new sub-field of research known as chiral waveguide QED~\cite{Lodahl2017}.
The chiral regime is naturally accessible within a nanophotonics platform, because the transverse confinement of light in optical nanowaveguides links the propagation direction of an emitted photon to the local polarization of an atom \cite{Lodahl2017,Bliokh2015}. This effect has been leveraged to achieve directional emission of optical photons in photonic waveguides and nanofibers \cite{Petersen2014,Mitsch2014,Sollner2015,Coles2016}.
However, to the best of our knowledge, directional emission of microwave photons into chiral waveguides for integration with circuit QED systems has not yet been demonstrated experimentally.}

In this work, we experimentally demonstrate on-demand directional photon emission based on the quantum interference of indistinguishable photons emitted by a giant artificial molecule.
We arrange qubits that are spatially separated along a bidirectional waveguide to form a giant artificial molecule that can 
\textcolor{black}{emit} photons in a chosen direction\textcolor{black}{~\cite{Gheeraert2020, Guimond2020, Solano2021, Redchenko2022}}. 
\textcolor{black}{Effectively, we create a chiral waveguide by linking the propagation direction of an emitted  photon to the relative phase of a two-qubit entangled state of the giant artificial molecule.} 
We use quadrature amplitude detection to obtain the moments of the two output fields of the waveguide.
Using these moments, we reconstruct the state of the photon and quantify its fidelity.
The architecture realized here can be used \textcolor{black}{for both photon emission and absorption~\cite{Gheeraert2020}, thus this demonstration is the first step towards implementing an interconnect for an extensible quantum network.}

\section{Experiment}
Our device comprises of four frequency-tunable transmon qubits~\cite{Koch2007} and four tunable transmon couplers~\cite{yan2018,sung2020} between each neighboring qubit pair, as shown in Fig.~\ref{fig:fig1}a.
The artificial molecule comprises two qubits $\textrm{Q}_1$ and $\textrm{Q}_2$, each of which resonantly emits photons with a frequency of $\omega_1/2\pi = \omega_2/2\pi = 4.93$ GHz, are equally coupled to a common waveguide with strength $\gamma/2\pi = 3.2$ MHz, and are spatially separated along the waveguide by a distance $\Delta x = \lambda/4$, where $\lambda$ is the wavelength of the emitted photon.
The remaining two qubits, $\textrm{Q}_3$ and $\textrm{Q}_4$, serve as data qubits that are not subject to direct dissipation into the waveguide.
These qubits would act as the interface between a quantum processor and the emitter qubits within a node.
The state of $\textrm{Q}_3$ and $\textrm{Q}_4$ can be prepared with high fidelity with a combination of single- and two-qubit gates.
Photons are generated by transferring the state of the data qubits $\textrm{Q}_{3/4}$ to the emitter qubits $\textrm{Q}_{1/2}$ via an exchange interaction mediated by the couplers $\textrm{C}_{13/24}$. 

\begin{figure*}[t!]
    \centering
    \includegraphics[width=\textwidth]{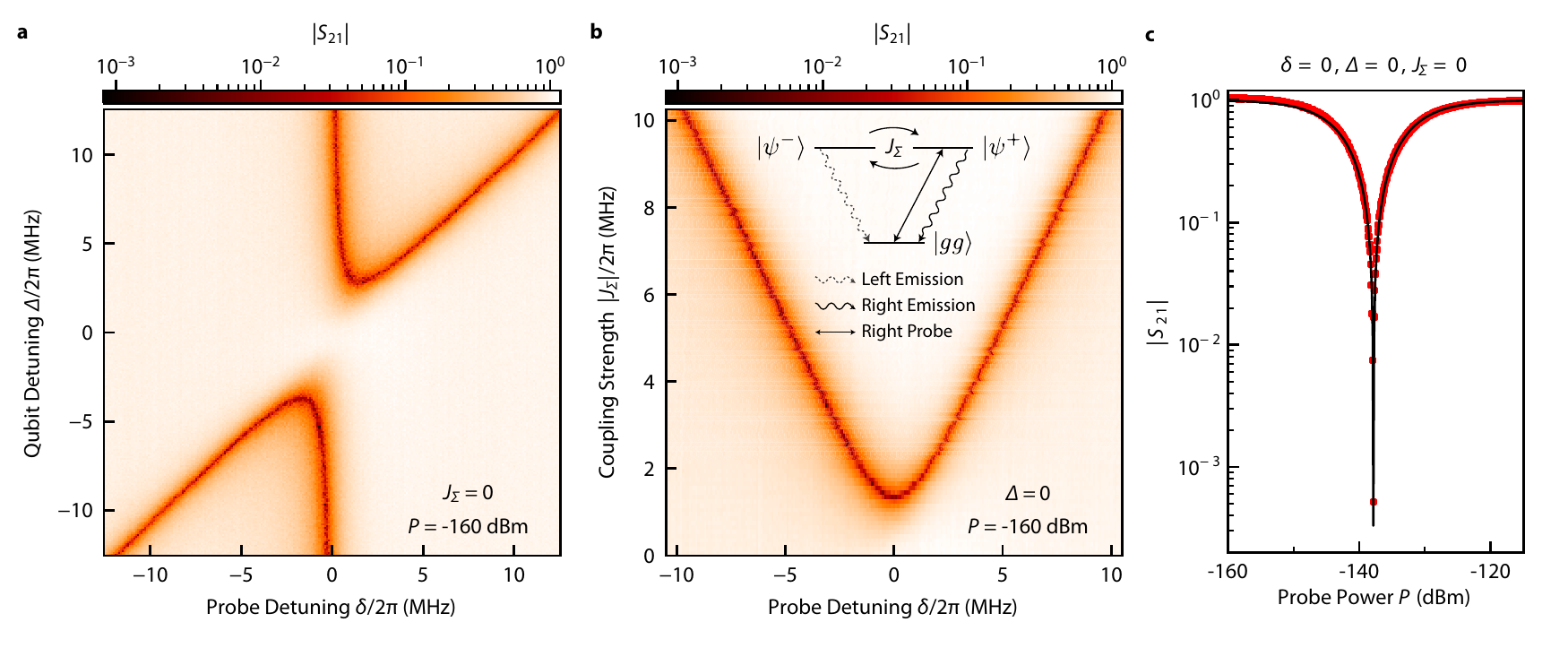}
    
    \caption{
    \textbf{Verifying protocol conditions via elastic scattering.}
    \textbf{a)} The transmittance $|S_{21}|$ of an input probe tone incident upon the two emitter qubits $\textrm{Q}_1$ and $\textrm{Q}_2$ through the waveguide. $|S_{21}|$ is plotted as a function $\Delta$, the detuning of $\textrm{Q}_2$ from $\textrm{Q}_1$, and $\delta$, the detuning between the probe and $\textrm{Q}_1$.
    When the qubits are far from resonance with each other ($\Delta > \gamma$), they will act as mirrors ($|S_{21}|\ll 1$) to the probe if the probe is resonant with either qubit ($\delta = 0, \Delta$). However, when the qubits are resonant ($\Delta = 0$), the transmittance returns to unity.
    \textbf{b)} $|S_{21}|$ as a function of the total coupling strength $|J_\Sigma|$ and $\delta$ while keeping $\textrm{Q}_1$ and $\textrm{Q}_2$ resonant and using the same probe power as in (a).
    The level diagram of the three states $|gg\rangle$, $|\psi^+\rangle$, and $|\psi^-\rangle$ is shown as an inset ($|ee\rangle$ is ignored for weak probes).
    The rightward-propagating probe used to obtain this data only couples the states $|gg\rangle\leftrightarrow|\psi^+\rangle$, and a finite exchange interaction between the emitters will couple $|\psi^+\rangle\leftrightarrow|\psi^-\rangle$.
    The state $|\psi^+\rangle$ can only emit a rightward-propagating photon and $|\psi^-\rangle$ can only emit a leftward-propagating photon.
    We observe two dips in the transmission at $\delta = \pm J_\Sigma$, corresponding to the energy splitting from the hybridization of $|\psi^\pm\rangle$.
    When $|J_\Sigma|\rightarrow0$, the transmission approaches unity for all $\delta$ because $|\psi^+\rangle$ is the only state that is excited, and it can only emit in the same direction (right) as the probe.
    This measurement is used to set $|J_\Sigma|=0$.
    \textbf{c)} The measured $|S_{21}|$ (red points) as a function of the probe power with $\Delta=0$, $\delta=0$, and $J_\Sigma = 0$.
    The data agree with a fit (black curve) to \textcolor{black}{a master equation simulation of the driven two-qubit system (see Supplementary \textcolor{red}{Fig. S3}).}}
    \label{fig:fig2}
\end{figure*}

\subsection{Protocol for Directional Emission}
The physics of the directional emission protocol is determined by the dynamics of the sub-system comprising the emitter qubits $\textrm{Q}_{1/2}$ and the waveguide.
For $\Delta x = \lambda/4$, the master equation that determines the time evolution of the emitters is \cite{Lalumiere2013,Gheeraert2020}:
\begin{equation}
    \partial_t \hat{\rho} = -i\big[\hat{H}_\textrm{qb} + \hat{H}_\textrm{c},\hat\rho\big] + \gamma \sum_j^2 D\big[\hat{\sigma}_j^-\big]\hat\rho,
    \label{eq: ME}
\end{equation}
where $\hat\rho$ is the density matrix of the sub-system, $D[\hat{O}] = \hat{O}\hat\rho\hat{O}^\dagger - \frac{1}{2} \{\hat{O}^\dagger\hat{O},\hat\rho\}$ is the Lindblad dissipator, $\hat{H}_\textrm{qb} = \sum_j^2 \omega_j \hat{\sigma}_j^+\hat{\sigma}_j^-$ is the bare Hamiltonian of the emitter qubits, and $\hat{\sigma}_j^\pm$ are the raising and lowering Pauli operators with $j\in\{1,2\}$.
Finally, $\hat{H}_\textrm{c} = (\gamma/2 + J_\textrm{c})(\hat\sigma_1^+\hat\sigma_2^- + \hat\sigma_2^+\hat\sigma_1^-)$ accounts for the exchange coupling between the qubits from two sources: a static waveguide-mediated interaction with strength $\gamma/2$ and a tunable-coupler-mediated interaction (via $\textrm{C}_{12}$) with strength $J_\textrm{c}$.
The tunability in $J_\textrm{c}$ is used to cancel the waveguide-mediated interaction such that the emitters are decoupled from each other. 

The final state of the photons emitted by $\textrm{Q}_1$ and $\textrm{Q}_2$ depends on the interference between their simultaneous emission.
Specifically, when the initial state of the emitter qubits is 
\begin{equation}
|\psi^{\pm}\rangle = \frac{|eg\rangle + e^{\pm i\frac{\pi}{2}}|ge\rangle}{\sqrt{2}},
\end{equation}
the node will emit a single photon that propagates either leftward or rightward, depending on the sign of the relative phase.
To see this, consider the emitter qubits initialized to $|\psi_\textrm{qb}\rangle=|\psi^{+}\rangle$, as shown in Fig.~\ref{fig:fig1}b.
There are four emission pathways from this state, each involving one of the emitter qubits, $\textrm{Q}_1$ or $\textrm{Q}_2$, releasing a photon that propagates towards the left or the right.
For simplicity, we define the positions of $\textrm{Q}_1$ and $\textrm{Q}_2$ along the waveguide to be $x=0$ and $x=\Delta x$, respectively. 
The pathways with a photon emitted by $\textrm{Q}_2$ will accumulate additional phases from both the relative phase $e^{i\pi/2}$ in $|\psi^+\rangle$ and a phase $e^{\pm ik\Delta x}$ from the position of $\textrm{Q}_2$ relative to $\textrm{Q}_1$.
Here, $k=2\pi/\lambda$ is the wavevector of the emitted photon, and the sign of the phase is determined by the propagation direction of the photon ($+$ for leftward, and $-$ for rightward).
These additional phases result in the total constructive (destructive) interference between the pathways that involve a photon propagating towards the right (left).
Therefore, the emitted photon solely propagates to the right in the state $|\psi_\textrm{ph}\rangle = |01\rangle$, where $|n_\textrm{L}n_\textrm{R}\rangle$ denotes the number of photons in the leftward- and rightward-propagating modes of the waveguide.
A similar analysis for the initial qubit state $|\psi_\textrm{qb}\rangle=|\psi^{-}\rangle$ is shown in Fig.~\ref{fig:fig1}c, indicating that the emitted photon propagates towards the left in the state $|\psi_\textrm{ph}\rangle = |10\rangle$ in this case.

The directional emission can be formally verified using the input-output relations for leftward- and rightward-propagating modes in the waveguide~\cite{Lalumiere2013}:
\begin{equation}
\label{eq:inout}
\begin{split}
        &\hat{a}_\textrm{L}^{} = \hat{a}_\textrm{L}^\textrm{in} + \sqrt{\frac{\gamma}{2}} (\hat{\sigma}_1^- + \hat{\sigma}_2^-e^{ik\Delta x}), \\
        &\hat{a}_\textrm{R}^{} = \hat{a}_\textrm{R}^\textrm{in} + \sqrt{\frac{\gamma}{2}} (\hat{\sigma}_1^- + \hat{\sigma}_2^-e^{-ik\Delta x}).
\end{split}
\end{equation}
Here, $\hat{a}_\textrm{L(R)}^\textrm{in}$ represents the input field of photons in the waveguide for the leftward (rightward) propagating mode.
From these relations, the number of photons in either mode of the waveguide, $\langle\hat{N}_\textrm{L(R)}\rangle = \langle \hat{a}_\textrm{L(R)}^\dagger\hat{a}_\textrm{L(R)}\rangle$, can be directly related to the state of the qubits.
Given that the emitters are initialized into one of $|\psi^\pm\rangle$, the interference described above is only perfect when $\Delta x = (2n+1) \lambda/4$, where $n$ is an integer, and $J_\textrm{c} = -\gamma/2$.
The first condition ensures that the interfering emission pathways are fully in/out of phase.
Additionally, it is the only spatial separation for which there is no correlated dissipation between the qubits~\cite{Lalumiere2013}, which would further disturb the interference.
The second condition prevents any population transfer between the qubits during the emission process by setting the exchange Hamiltonian $\hat{H}_\textrm{c}$ to zero. 

\begin{figure}[t!]
    \centering
    \includegraphics[width=3.45in]{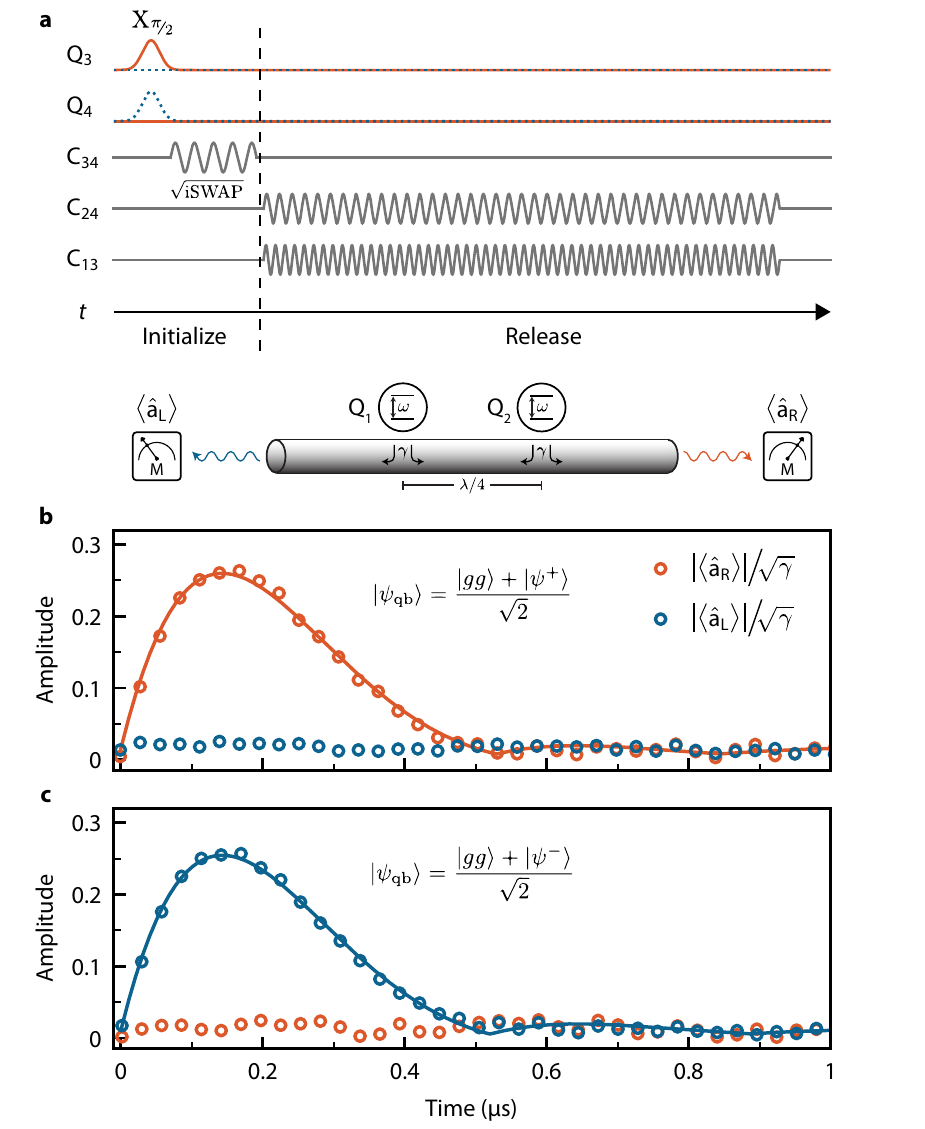}
    \caption{
    \textbf{Pulse sequence and time-domain measurements.}
    \textbf{a)} The pulse sequence for generating a photon.
    The qubit state initialization begins by exciting either $\textrm{Q}_3$ (orange solid curve) or $\textrm{Q}_4$ (blue dashed curve).
    These qubits are then entangled via a $\sqrt{\textrm{iSWAP}}$ gate by parametrically modulating the frequency of the tunable coupler $\textrm{C}_{34}$ at the detuning of $\textrm{Q}_3$ and $\textrm{Q}_4$.
    Finally, a photon is released via a parametric exchange interaction between the data qubits $\textrm{Q}_{3/4}$ and the emitter qubits $\textrm{Q}_{1/2}$.
    The measurement schematic below the pulse-sequence shows that the field amplitudes $\hat{a}_\textrm{L/R}$ are acquired at both outputs of the waveguide.
    \textbf{b)} The measured (circles) time-dependent field amplitudes for a photon emitted towards the right.
    The data is fit (solid curve) using the solution to the master equation (see Supplementary Info.).
    The initial state of the data qubits is $|\psi_\textrm{qb}\rangle = (|gg\rangle + |\psi^+\rangle)/\sqrt{2}$.
    The field amplitude of the leftward emission channel is nearly zero. \textcolor{black}{This data is averaged over $1.5 \times 10^7$ repetitions.}
    \textbf{c)} The same measurement as (b), but with $|\psi_\textrm{qb}\rangle = (|gg\rangle + |\psi^-\rangle)/\sqrt{2}$ such that the emitted photon now propagates to the left.}
    \label{fig:fig3}
\end{figure}

\subsection{Device Calibration for Directional Emission}
Verifying that the ideal directional emission conditions are satisfied in the experiment is challenging.
In particular, the strong and always-on dissipation into the waveguide makes it difficult to measure the strength of the coupling between the emitters, $J_\Sigma = \gamma/2 + J_\textrm{c}$.
The typical methods, such as observations of avoided crossings in qubit spectroscopy or population exchange in time domain, are limited in resolution when outside the strong coupling regime where $J_\Sigma < \gamma$.
To go beyond this limit, we infer the value of $J_\Sigma$ by measuring the elastic scattering of a weak input probe tone.
Specifically, we measure the transmission amplitude $S_{21}$ of a coherent tone as a function of the detuning between the emitter qubit frequencies, $\Delta = \omega_2 - \omega_1$, and the detuning between the probe and $\textrm{Q}_1$ frequencies, $\delta = \omega_p - \omega_1$, as shown in Fig.~\ref{fig:fig2}a.
When the qubits are detuned ($\Delta > \gamma$), they will each act as a mirror to single photons at their respective frequencies~\cite{Astafiev2010,Hoi2011,Mirhosseini2019}, such that there are two dips in $|S_{21}(\delta)|$.
This is a consequence of the destructive interference between the probe and the forward-propagating, out-of-phase emission of the driven qubit.
Therefore, $|S_{21}|$ is suppressed for weak coherent inputs (average photon number $\ll1$) that are resonant with either qubit.

The elastic scattering behavior changes when the emitter qubits are resonant ($\Delta = 0$).
First, given that the qubits are equally coupled to the waveguide, the input probe tone will only drive the $|gg\rangle \leftrightarrow |\psi(\phi)\rangle$ and $|\psi(\phi)\rangle \leftrightarrow |ee\rangle$ transitions, where $|\psi(\phi)\rangle = (|eg\rangle + e^{i\phi}|ge\rangle)/\sqrt{2}$.
The sign of $\phi = \pm k\Delta x$ is determined by the propagation direction of the probe.
Furthermore, the second transition can be ignored for low probe powers $P$, as it requires an appreciable population in $|\psi(\phi)\rangle$ to play a role.
Therefore, if $\Delta x = \lambda/4$ and $\hat{H}_\textrm{c} = 0$, the state of the qubits will be driven into a mixture of only $|gg\rangle$ and one of $|\psi^+\rangle$ or $|\psi^-\rangle$, depending on the direction of the probe.
However, these states can only re-emit photons in the same direction as the input, as depicted in the level-diagram in Fig.~\ref{fig:fig2}b for a rightward-propagating probe. This \textcolor{black}{ideally} results in perfect transmission, $|S_{21}(\Delta = 0)|=1$. 

The magnitude of the transmission will deviate from unity if $\hat{H}_\textrm{c} \neq 0$, as any population transfer between $|\psi^+\rangle\leftrightarrow|\psi^-\rangle$ will cause part of the qubit emission to propagate in the  direction opposite to that of the probe.
To verify this, we measure $|S_{21}(\Delta = 0)|$ as a function of $|J_\Sigma|$ in Fig.~\ref{fig:fig2}b.
For $|J_\Sigma| > \gamma/2$ we see two dips in the transmission at $\delta = \pm J_\Sigma$, which now correspond to the hybridized energy splitting of $|\psi^+\rangle$ and $|\psi^-\rangle$.
For $|J_\Sigma| < \gamma/2$, the energy splitting is within the linewidth of the qubits, which is set by $\gamma$.
However, as described above, we observe the $|S_{21}(\delta)|\rightarrow1$ as $J_\Sigma\rightarrow0$.
Therefore, we can use the transmission as a metric to set $J_\Sigma=0$ despite the large decay rate $\gamma$ of these qubits. 

Finally, in Fig.~\ref{fig:fig2}c we show the transmission $|S_{21}(\Delta = 0, \delta = 0, J_\Sigma = 0)|$ as a function of the probe power.
Here, we clearly see $|S_{21}| \textcolor{black}{\rightarrow} 1$ for both low powers, as previously discussed, and high powers, where the average photon number of the probe is much greater than one and the emitter qubits are fully saturated.
For intermediate powers, however, the transmission is no longer unity, because the qubits are neither fully saturated nor restricted to the zero- and single-excitation subspace.
That is, the population of $|ee\rangle$ and its subsequent decay into both $|\psi^\pm \rangle$ cannot be ignored, in contrast to the simpler low-power case.
We numerically simulate the power-dependence of the transmission amplitude using input-output theory. \textcolor{black}{For low powers, we observe that $|S_{21}|$ slightly exceeds unity, which we attribute to impedance mismatches in our experimental setup~\cite{Khalil2012,Probst2015}.
Apart from this,} the resulting simulation fits well to the data in Fig.~\ref{fig:fig2}c, demonstrating the validity of our model.
\textcolor{black}{The power dependence of the transmission is similar to that of the reflection of a single emitter coupled to a semi-infinite waveguide~\cite{Hoi2015, Scigliuzzo2020}.
In this sense, two qubits coupled to a bidirectional chiral waveguide 
resembles a single qubit coupled to a semi-infinite waveguide.}

\subsection{Photon Generation and Measurement}
Having realized the conditions required to observe directional photon emission, we now run the full protocol using the pulse sequence shown in Fig.~\ref{fig:fig3}a.
Rather than directly preparing the initial state of the emitter qubits into $|\psi^\pm\rangle$, which have low coherence due to their continuous dissipation into the waveguide, we instead initialize qubits $\textrm{Q}_3$ and $\textrm{Q}_4$, which have longer lifetimes.
We do so by first exciting either $\textrm{Q}_3$ or $\textrm{Q}_4$ while they are decoupled.
Next, the frequency of the tunable coupler $\textrm{C}_{34}$ is modulated at the detuning of this qubit pair to implement an entangling $\sqrt{\mathrm{iSWAP}}$ gate \cite{McKay2016}.
Depending on which qubit was initially excited, the $\sqrt{\mathrm{iSWAP}}$ gate will take the combined state of $\textrm{Q}_3$ and $\textrm{Q}_4$ to one of $|\psi^\pm\rangle$.
Parametric exchange interactions mediated by tunable couplers $\textrm{C}_{13}$ and $\textrm{C}_{24}$ are used to transfer the state of $\textrm{Q}_3$ and $\textrm{Q}_4$ into $\textrm{Q}_1$ and $\textrm{Q}_2$ (see Supplementary \textcolor{red}{Fig. S5}.), which simultaneously emit their excitations as photons.
The interference process in Fig.~\ref{fig:fig1} remains the same, but the shape of the emitted photon is now determined by both the parametrically induced coupling $g_\mathrm{eff}$ between the qubit pairs $\textrm{Q}_{1/2} \leftrightarrow \textrm{Q}_{3/4}$ and $\gamma$.

\begin{figure*}[t!]
    \centering
    \includegraphics[width=\textwidth]{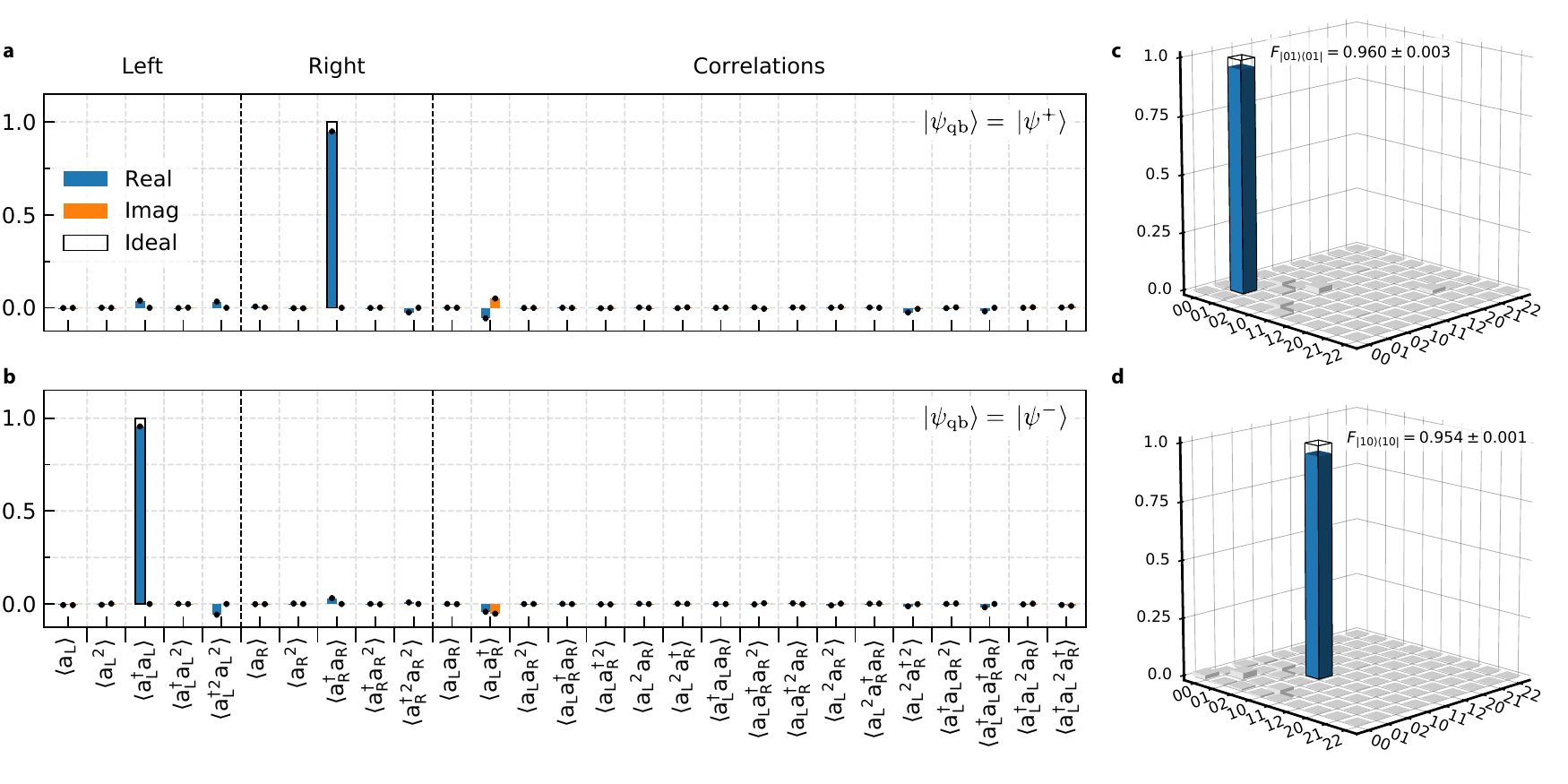}
    
    \caption{
    \textcolor{black}{\textbf{Photon state tomography. a)}
    The moments and correlations of the left and right propagating channels of the waveguide up to $4^\mathrm{th}$ order with $|\psi_\textrm{qb}\rangle = |\psi^+\rangle$.
    All moments are nearly zero, except $\langle\hat{a}^\dagger_\textrm{R}\hat{a}_\textrm{R}\rangle \approx 0.95$. \textcolor{black}{This data is averaged over $5 \times 10^8$ repetitions.}
    \textbf{b)} The same as (a) but with $|\psi_\textrm{qb}\rangle = |\psi^-\rangle$.
    All moments are once again nearly zero, except $\langle\hat{a}^\dagger_\textrm{L}\hat{a}_\textrm{L}\rangle \approx 0.95$.
    \textbf{c)} The real part of the density matrix of the photon emitted to the right based on the moments shown in (a) with a state fidelity of $F_{|01\rangle\langle01|}=0.96\pm 0.003$. The Hilbert space of the emitted photon is truncated to $N\leq2$ photons.
    \textbf{d)} The real part of the density matrix of the photon emitted to the left based on the moments shown in (b) with a state fidelity of $F_{|10\rangle\langle10|}=0.954\pm 0.001$.}} 
    \label{fig:fig4}
\end{figure*}

We first measure the temporal dynamics of the averaged field amplitudes $\hat{a}_\textrm{L/R}(t)$.
The field amplitudes are only non-zero when there is finite coherence between the $|00\rangle$ and $|01\rangle$ or $|10\rangle$ states.
Indeed, if $\textrm{Q}_3$ and $\textrm{Q}_4$ are initialized in the state $|\psi ^\pm\rangle$, such that the emitted photon is in a Fock state, the field amplitude will be zero.
Therefore, we initially excite $\textrm{Q}_\text{3}$ ($\textrm{Q}_\text{4}$) with a $\frac{\pi}{2}$-pulse, such that the emitted photon will be in the state \textcolor{black}{with maximal coherence,} $[|00\rangle + |01\rangle]/\sqrt{2}$ ($[|00\rangle + |10\rangle]/\sqrt{2}$).
The photon wavepacket is now visible \textcolor{black}{with maximized field amplitude}, as shown in Figs.~\ref{fig:fig3}b and~\ref{fig:fig3}c.
The amplitude of the photon is non-zero in only a single direction that is determined by the phase in the initial state of $\textrm{Q}_3$ and $\textrm{Q}_4$, a signature of the controlled directional emission.
We fit this data (see Supplementary \textcolor{red}{Fig. S6}) to obtain the effective coupling between the data and emitter qubit pairs $g_\mathrm{eff}/2\pi = 1.28$ MHz.

Next, we perform photon state tomography~\cite{Eichler2011,Eichler2012,Lang2013,kannanN00N2020} to fully reconstruct the state of the emitted photon and quantify its fidelity.
We use quadrature amplitude detection of the left and right outputs of the waveguide to obtain the higher-order moments and correlations of the fields.
Time-independent values of the field quadratures $S_\textrm{L/R} = X_\textrm{L/R} + iP_\textrm{L/R}$ are obtained by digitally demodulating and integrating individual records of the measured time-dependent field amplitudes.
Using repeated measurements of these values, we construct a 4D probability distribution $D(S_\textrm{L},S_\textrm{L}^*,S_\textrm{R},S_\textrm{R}^*)$ that is used to obtain the moments of $S_\textrm{L}$ and $S_\textrm{R}$,
\begin{equation}
\label{eq:moments}
\begin{split}
&\langle {\hat{S}_\textrm{L}^{\dag w}} \hat{S}_\textrm{L}^x {\hat{S}_\textrm{R}^{\dag y}} \hat{S}_\textrm{R}^z \rangle = \\ &\int d^2S_\textrm{L} d^2S_\textrm{R}\ S_\textrm{L}^{*w} S_\textrm{L}^x S_\textrm{R}^{*y} S_\textrm{R}^z\ D(S_\textrm{L},S_\textrm{L}^*,S_\textrm{R},S_\textrm{R}^*),
\end{split}
\end{equation}
where $w,x,y,z \in \{0,1,2,...\}$.
The measured signals $S_\textrm{L/R}$ are composed of both the field of interest $\hat a_\textrm{L/R}$ as well as noise added by the amplification chain.
This additional noise is subtracted from the moments of $\hat{S}_\textrm{L/R}$, using the input-output relations for phase-insensitive amplifiers~\cite{Caves1982}, to obtain the desired moments of $\hat a_\textrm{L/R}$~\cite{Eichler2011,Eichler2012, kannanN00N2020}. \textcolor{black}{These moments are normalized by the gain of the amplification chain from the qubits to the electronics used for signal acquisition.}

The moments of and correlations between $\hat{a}_\textrm{L}$ and $\hat{a}_\textrm{R}$ for the photons we generate are shown in Figs.~\ref{fig:fig4}a and~\ref{fig:fig4}b \textcolor{black}{up to fourth order}. When $\textrm{Q}_3$ and $\textrm{Q}_4$ are initialized to $|\psi^+\rangle$, we obtain $\langle \hat a ^\dagger_\textrm{R} \hat a_\textrm{R} \rangle \approx 1$ as the only appreciably non-zero moment, as expected for a single photon which only propagates towards the right.
Similarly, we measure $\langle \hat a ^\dagger_\textrm{L} \hat a_\textrm{L} \rangle \approx 1$ as the only non-zero moment for the leftward-propagating photon emitted when the qubits are initialized to $|\psi^-\rangle$. \textcolor{black}{All third and fourth order moments are nearly zero (maximum magnitude of 0.05), demonstrating the single-photon nature of the emission process.}

Finally, we use these moments to obtain the density matrices of the emitted photons, shown in Figs.~\ref{fig:fig4}c and~\ref{fig:fig4}d, using maximum-likelihood-estimation \textcolor{black}{\cite{Chow2012,Eichler2012}}. \textcolor{black}{Here, we truncate the Hilbert space to $N\leq2$ photons.}
From these density matrices, we obtain a state fidelity of \textcolor{black}{$F = 0.960 \pm 0.003$} and \textcolor{black}{$F = 0.954 \pm 0.001$} for the rightward- and leftward-propagating photons, respectively.
We observe a small, non-zero number of photons in the right (left) output of the waveguide when the qubits are initialized to $|\psi^+\rangle$ ($|\psi^-\rangle$).
This infidelity results from imperfect interference between the emission pathways caused by qubit decoherence during emission and small deviations from necessary conditions $\Delta x = \lambda/4$ and $J_\Sigma=0$.

\section{Discussion}
Our results demonstrate that quantum interference between emitters in a waveguide QED architecture can be used to realize a directional single photon source.
While we have only performed photon generation in this work, the time-reverse of the emission protocol can be used to capture photons with this same architecture if the wavepacket of the incoming photon is symmetric in time~\textcolor{black}{\cite{Kurpiers2018, Axline2018, Campagne-Ibarcq2018, Gheeraert2020}}.
Note that the wavepacket of the generated photon can be shaped arbitrarily, in principle, by varying the time-dependence of the coupling between the data and emitter qubits\textcolor{black}{~\cite{Yin2013, Pechal2014, Forn-Diaz2017,Kurpiers2018, Axline2018, Campagne-Ibarcq2018, Gheeraert2020,reuer2021realization}}.
Looking forward, we envision building a quantum network by tiling devices with the presented architecture in series and applying our protocol for both photon generation and capture. \textcolor{black}{Error mitigation strategies compatible with this architecture include heralding, entanglement purification \cite{Yan2022}, teleportation with GHZ states \cite{Greenberger1990}, and quantum communication with W states \cite{Dur2001}.}
Such a network will enable entanglement distribution and information shuttling with high fidelity in support of extensible quantum information processing.

\section*{Author Contributions}
B.K. designed the experiment procedure.
B.K. and A.A. designed the devices, conducted the measurements, analyzed the data, and wrote the manuscript. A.D.P. provided theory support.
A.M. and B.M.N. performed sample fabrication.
Y.S., D.A.R., K.S., and J.I-J.W. assisted with the experimental setup.
R.W. developed the custom FPGA code used to obtain the data.
J.B., A.K., and A.V. assisted with the automation of the device calibration.
M.E.S., J.L.Y, T.P.O., S.G., J.A.G., and W.D.O. supervised the project.
All authors discussed the results and commented on the manuscript. 
\section*{Acknowledgments}
The authors gratefully acknowledge Daniel Campbell for his contributions to the infrastructures used in this experiment, and David K. Kim for assisting with device fabrication.
This research was funded in part by the AWS Center for Quantum Computing, U.S. Army Research Office Grant No. W911NF-18-1-0411, the DOE Office of Science National Quantum Information Science Research Centers, Co-design Center for Quantum Advantage (C2QA) under Contract No. DE-SC0012704, and the Department of Defense under Air Force Contract No. FA8702-15-D-0001.
B.K. gratefully acknowledges support from the National Defense Science and Engineering Graduate Fellowship program.
A.A. gratefully acknowledges support from the P.D. Soros Fellowship program.
Any opinions, findings, conclusions or recommendations expressed in this material are those of the author(s) and should not be interpreted as necessarily representing the official policies or endorsements of the U.S. Government.

\section*{Data Availability}
The data that support the findings of this study are available from the corresponding author upon reasonable request.

\section*{Code Availability}
The code used for numerical simulations and data analyses is available from the corresponding author upon reasonable request.

\bibliography{main}

\onecolumngrid
\newpage
\begin{center}
    \textbf{SUPPLEMENTARY INFORMATION}
\end{center}

\setcounter{figure}{0}
\setcounter{equation}{0}
\makeatletter 
\renewcommand{\thefigure}{S\@arabic\c@figure}
\renewcommand{\thetable}{S\@arabic\c@table}
\renewcommand{\theequation}{S\arabic{equation}}
\makeatother
\begin{center}
    \textbf{Device and Experimental Setup}
\end{center}
\newcolumntype{P}[1]{>{\centering\arraybackslash}p{#1}}

\begin{center}

\end{center}

\begin{figure*}[h]
    \centering
    \includegraphics[width=\textwidth]{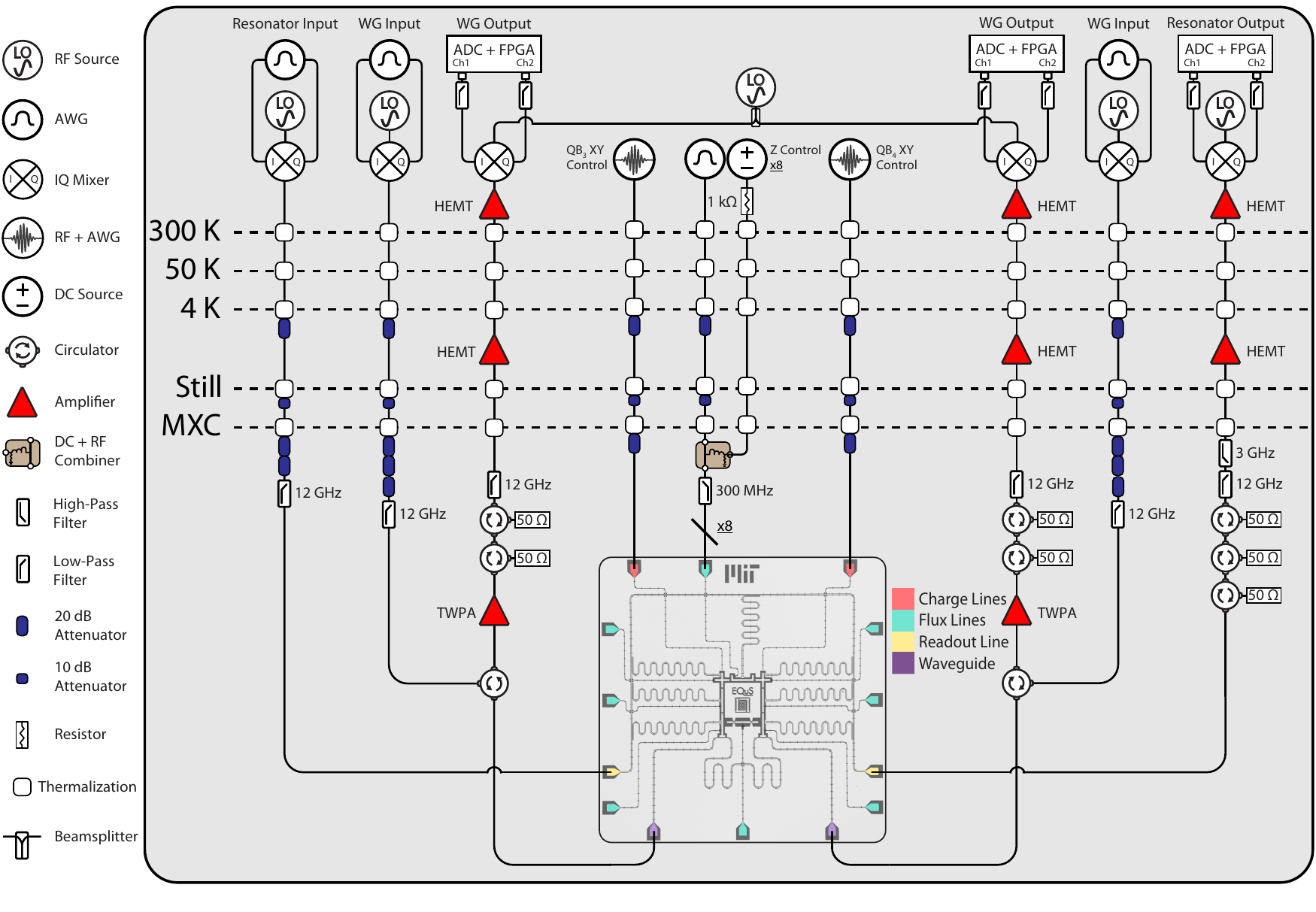}
    
    \caption{\textbf{Experimental setup.} Wiring schematic of the device and all electronics used to perform the experiment. Note that only one flux line configuration is shown (green), but each qubit and coupler is coupled to a flux line with separate, but identical, control electronics.}
    \label{fig:setup}
\end{figure*} 

This experiment was conducted in a Bluefors XLD1000 dilution refrigerator, which can reach a base temperature of 10 mK. The experimental setup is shown in Fig.~\ref{fig:setup}. The device is protected from ambient magnetic fields by superconducting and Cryoperm-10 shields below the mixing chamber (MXC). Each end of the waveguide is connected to a microwave circulator for dual input-output operation. To minimize thermal noise from higher temperature stages, the inputs are attenuated by 20 dB at the 4K stage, 10 dB at the 1K stage, and 60 dB (40 dB for resonator readout input) at the MXC. The output signals are filtered with $\SI{3}{GHz}$ high-pass and $\SI{12}{GHz}$ low-pass filters. Two additional isolators are placed after the circulator in the MXC to prevent noise from higher-temperature stages travelling back into the sample. High electron mobility transistor (HEMT) amplifiers are used at $\SI{4}{K}$ and room-temperature stages of the measurement chain to amplify the outputs from the device. The signals are then downconverted to an intermediate frequency using an IQ mixer, filtered, digitized, and demodulated. All qubits and tunable couplers are also equipped with their own flux bias lines. A DC + RF combiner is used for all flux lines to provide both static and dynamic control of the qubit/coupler frequencies. The DC and RF inputs are joined by a RF choke below the MXC before passing through a 300 MHz low pass filter. The RF flux control lines are attenuated by 20 dB at the $\SI{4}{K}$ stage, and by 10 dB at the 1K stage. The data qubits are equipped with local charge lines for independent single-qubit XY gates. The specific control and measurement equipment used throughout the experiment is summarized in Table \ref{tab:equipment}. The relevant parameters of the device used in the experiment are summarized in Table  \ref{tab:params}.

\begin{table}[h!]
\centering
\begin{tabular}{|c|c|c|}
\hline
Component       & Manufacturer     & Model   \\ \hline
Dilution Fridge & Bluefors         & XLD1000 \\
RF Source       & Rohde \& Schwarz & SGS100  \\
DC Source       & QDevil           & QDAC    \\
Control Chassis & Keysight         & M9019A  \\
AWG             & Keysight         & M3202A  \\
ADC             & Keysight         & M3102A  \\ \hline
\end{tabular}
\caption{\textbf{Summary of control equipment.} The manufacturers and model numbers of the control equipment used for the experiment.}
\label{tab:equipment}
\end{table}

    \begin{table}[h!]
        \centering
    \begin{tabular}{ P{2.5cm}|P{2cm}|P{2cm}|P{2cm}|P{2cm}}
 
         Parameter & $\textrm{Q}_1$ & $\textrm{Q}_2$ & $\textrm{Q}_3$ & $\textrm{Q}_4$ \\
         \hline
         Frequency & 4.93 GHz   & 4.93 GHz    &  4.8 GHz    & 4.85 GHz\\
         Anharmonicity  & -274 MHz    &-273 MHz      &  -307 MHz     &-307 MHz\\
         $\gamma/2\pi$ & 3.2 MHz & 3.2 MHz &  - & - \\
         $\gamma_\phi/2\pi$ & 8 kHz & 41 kHz & - & -\\   
         $T_1$    & - & - &  13.8 $\mu$s & $13.4\mu$s \\
         $T_2^*$ & - & - & 18.1 $\mu$s & 23.6 $\mu$s\\
        \end{tabular}
     \caption{\textbf{Summary of device parameters.} The operational qubit frequencies, anharmonicities, emitter-waveguide coupling strengths $\gamma$, emitter dephasing rates $\gamma_\phi$, and $T_1$ and $T_2^*$ of the data qubits are given for the emitter ($\textrm{Q}_{1/2}$) and data qubits ($\textrm{Q}_{3/4}$) on the device used throughout the experiment.}
     \label{tab:params}
    \end{table}

\begin{center}
    \textbf{Spectroscopic Measurements}
\end{center}

\begin{center}
    {I. Single-Qubit Scattering}
\end{center}

\begin{figure*}[t!]
    \centering
    \includegraphics[width=7in]{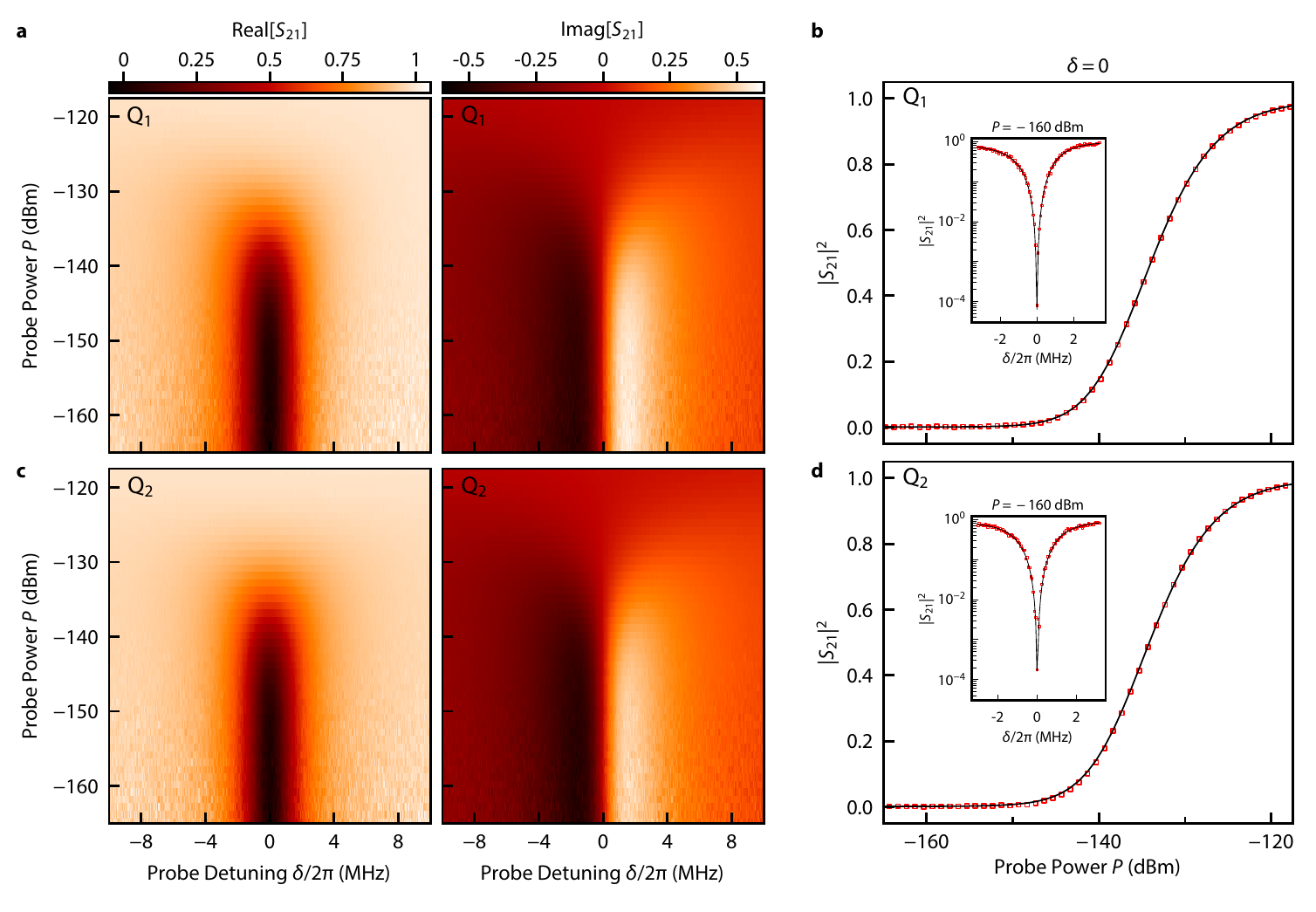}
    
    \caption{\textbf{Emitter qubit spectroscopy.} a, c) Real (left) and imaginary (right) components of the transmission spectrum of a coherent probe incident on Q$_1$ (Q$_2$) through the waveguide as a function of the qubit-probe detuning $\delta/2\pi$ and the probe power $P$. We extract the qubit-waveguide coupling rate $\gamma/2\pi = 3.2$ MHz and the qubit dephasing rate $\gamma_{\phi}/2\pi = 8$  kHz (41 kHz). b, d) Transmittance $|S_{21}|^2$ as a function of probe power P at zero qubit-probe detuning ($\delta/2\pi = 0$). The measured data is plotted in red, and the theoretical fit is plotted in black. The inset shows the frequency response of the emitter qubit at probe power $P$ = -160 dBm.}
    \label{fig:spec}
\end{figure*}

We measure the elastic scattering of a coherent input into the waveguide in order to extract device parameters that involve the emitter qubits. Consider a single emitter qubit that is strongly coupled to the waveguide with strength $\gamma$. The qubit will scatter coherent probe tones sent through the waveguide in a manner such that it acts as a mirror to single photons. Therefore, the scattering parameters will strongly depend on the probe power, since this is what determines the number of photons incident upon the emitter~\cite{Hoi2011, Astafiev2010, Hoi2013}. The master equation for the simplified model of a single emitter coupled to a waveguide is given by~\cite{Mirhosseini2019}
\begin{equation}
    \partial_t \hat{\rho} = -i\big[\hat{H},\hat\rho\big] + \gamma D\big[\hat{\sigma}^-\big]\hat\rho + \frac{\gamma_{\phi}}{2}D\big[\hat{\sigma}_z\big]\hat \rho.
\end{equation}
The single-emitter Hamiltonian is $\hat H = \frac{1}{2}\delta \hat\sigma_z  + \frac{1}{2}\Omega_p\hat\sigma_x$, $\gamma_{\phi}$ is the pure dephasing rate of the emitter, $\delta = \omega - \omega_p$ is the emitter-probe detuning, and $\Omega_p = \sqrt{2\gamma P/\hbar\omega_\mathrm{p}}$ is the drive strength of the probe with power $P$. Assuming that the probe propagates towards the right, the right-ward propagating output of the waveguide can be determined via input-output theory:
\begin{equation}
    \hat a_\mathrm{R} = \hat a_\mathrm{R}^\textrm{in} + \sqrt{\frac{\gamma}{2}}\hat \sigma^{-}.
\end{equation}
Therefore, the transmission amplitude $S_{21} = \langle\hat a_\mathrm{R}\rangle/\langle\hat a_\mathrm{R}^\textrm{in}\rangle$  can be calculated to be~\cite{Mirhosseini2019}
\begin{equation}
    S_{21}(\delta, \Omega_p) = 1 - \frac{\gamma(1 - i\frac{\delta}{\gamma_2})}{2\gamma_2\left(1 + \left(\frac{\delta}{\gamma_2}\right)^2 + \frac{\Omega_p^2}{\gamma\gamma_2} \right)},
\end{equation}
where $\gamma_2 = \gamma/2 + \gamma_{\phi}$ is the total decoherence rate of the emitter. Transmission measurements as a function of probe power $P$ and detuning $\delta$, as shown shown in Fig.~\ref{fig:spec}, allow us to obtain the fit parameters such as $\gamma/2\pi  \approx 3.2$ MHz and $\gamma_{\phi}/2\pi\approx  8$ kHz (41 kHz). These measurements serve as a method to calibrate the absolute power of microwave tones incident on the emitter qubits.

 \begin{figure}[t!]
     \centering
     \includegraphics[width = \textwidth]{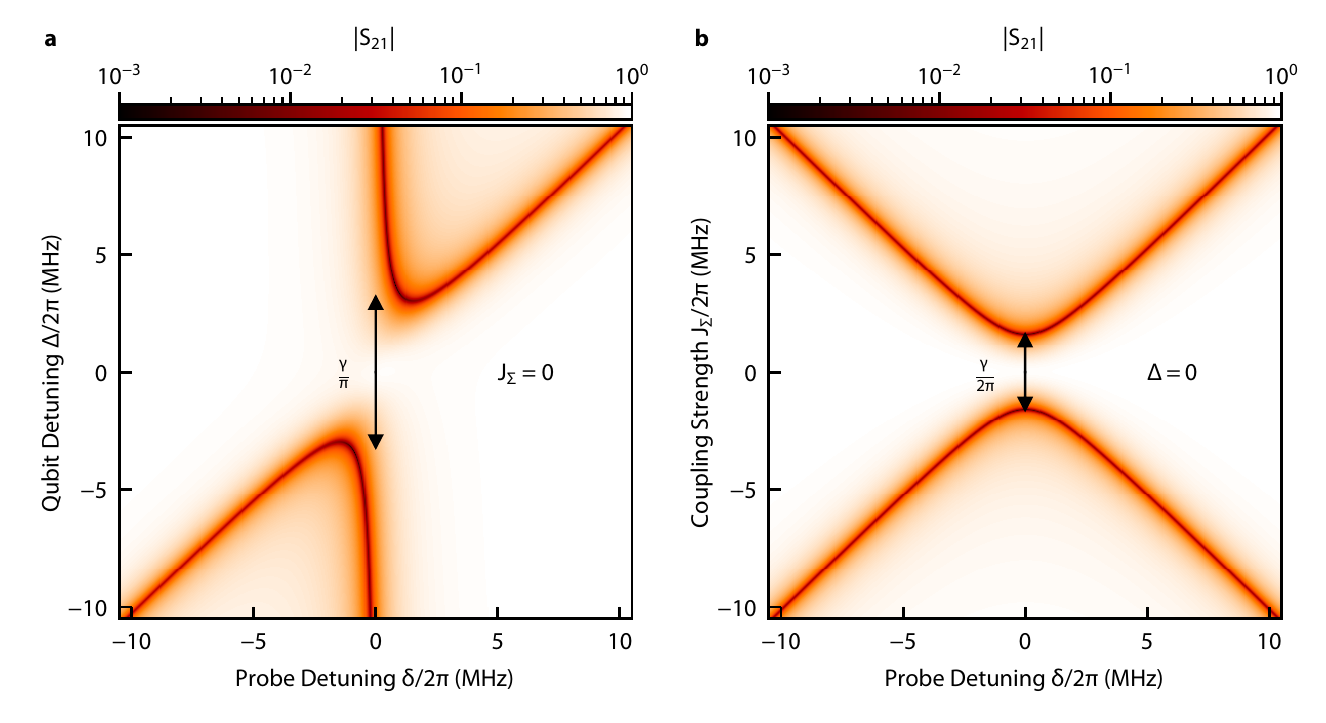}
     \caption{\textbf{Master equation simulations for emitter spectroscopy. a)} Simulated transmittance $|S_{21}|$ of an input probe tone of power $P = -160$ dBm incident upon the two uncoupled emitter qubits $\textrm{Q}_1$ and $\textrm{Q}_2$ through the waveguide ($J_\Sigma = 0$). $|S_{21}|$ is plotted as a function $\Delta$, the detuning of $\textrm{Q}_2$ from $\textrm{Q}_1$, and $\delta$, the detuning between the probe and $\textrm{Q}_1$. The model includes emitter-waveguide coupling strengths $\gamma/2\pi = 3.2$ MHz and dephasing rates $\gamma_{\phi, 1}/2\pi = 8$ kHz and $\gamma_{\phi, 2}/2\pi = 41$ kHz. \textbf{b)} Simulated $|S_{21}|$ of the same coherent probe incident on the emitter qubits through the waveguide as a function of the total coupling strength $|J_\Sigma|$ and probe detuning $\delta$ for resonant emitter qubits ($\Delta = 0$).}
     \label{fig:fig2sim}
 \end{figure}
 
  \begin{figure}[h!]
     \centering
     \includegraphics[width = \textwidth]{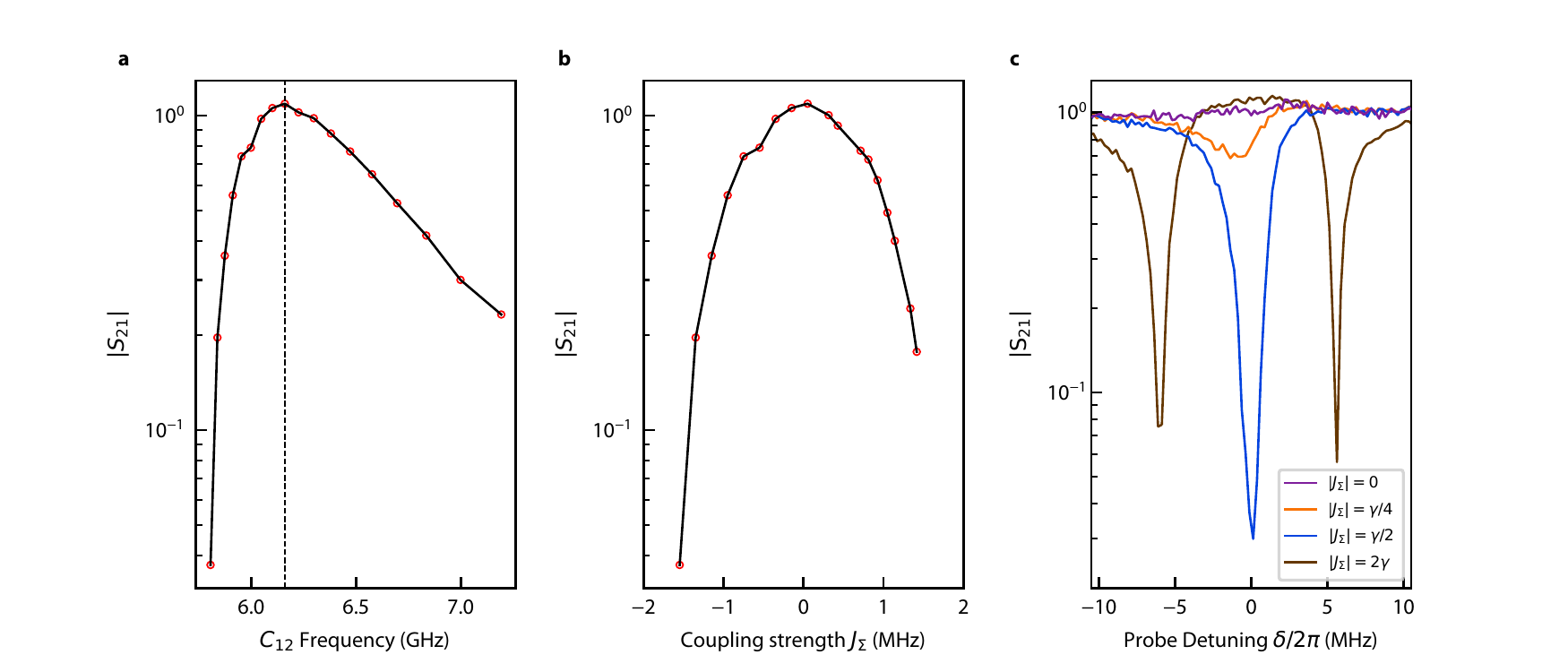}
     \caption{\textbf{Cancelling coupling between emitter qubits. a)} The transmission $|S_{21}|$ as a function of the frequency of the tunable coupler $C_{12}$ for resonant emitters. The coupling between the two emitter qubits is tuned via the frequency $C_{12}$, and the point of highest $|S_{21}|$ corresponds to the point of operation for a net-zero interaction between these qubits. \textbf{b)} The same data presented in (a), but with the frequency of $C_{12}$ mapped onto the net coupling strength $J_\Sigma$. For the range $J_\Sigma \in \left[-\gamma/2,\gamma/2\right]$, $|S_{21}|$ is maximal at $J_\Sigma=0$. \textbf{c)} Four representative traces of $|S_{21}|$ as a function of the detuning $\delta$ between the probe and resonant emitter qubits. A clear splitting splitting can be seen for $|J_\Sigma|=2\gamma$. The $|S_{21}|$ approaches unity for all $\delta$ as $|J_\Sigma|$ approaches 0.}
     \label{fig:cancel}
 \end{figure}
 
 \begin{center}
    {II. Two-Qubit Scattering}
\end{center}
 
 We extend the model to include both emitter qubits Q$_1$ and Q$_2$ of frequencies $\omega_1$ and $\omega_2$ at positions $x_1$ and $x_2$ along the waveguide.
 In order to calibrate the device, we perform a series of elastic scattering measurements discussed in the main text in Fig.~\ref{fig:fig2}.
 In the frame of the probe, the system Hamiltonian is \cite{Lalumiere2013}
 
 \begin{equation}
     \hat H = \hat H_\mathrm{qb} + \hat H_\mathrm{p} + \hat H_\mathrm{c} = \sum_j^2 [\delta_j \hat{\sigma}_j^+\hat{\sigma}_j^- +\Omega_\mathrm{p}(\hat \sigma_j^+e^{-ikx_j} + \hat \sigma_j^-e^{ikx_j})] +  J_{\Sigma}(\hat\sigma_1^+\hat\sigma_2^- + \hat\sigma_2^+\hat\sigma_1^-) 
 \end{equation}

where $\delta_j = \omega_j - \omega_\mathrm{p}$ is the probe detuning from each qubit frequency.
The total emitter-emitter coupling is $J_\Sigma = \gamma/2 + J_\mathrm{c}$, where $J_\mathrm{c}$ is the coupling induced by tunable coupler C$_{12}$ as discussed in the main text.
We define the positions of the emitters as $x_1 = 0$ and $x_2 = \lambda/4$, where $\lambda$ is the wavelength of the qubit emission in the waveguide.
The master equation of the driven qubit system is
\begin{equation}
    \partial_t \hat{\rho} = -i\big[\hat{H}, \hat\rho\big] + \sum_j^2\left(\gamma D\big[\hat{\sigma}_j^-\big]\hat\rho + \frac{\gamma_{\phi,j}}{2}D\big[\hat{\sigma}_{z,j}\big]\hat \rho \right).
    \label{eq:ME_ss}
\end{equation}

For a rightward-propagating input probe with average field amplitude $\langle \hat a_\mathrm{R}^\mathrm{in} \rangle = \sqrt{\frac{P}{\hbar \omega_\mathrm{p}}}$, the input-output relations of the driven two-qubit system are
\begin{equation}
\begin{split}
        &\langle\hat{a}_\textrm{L}\rangle =  \sqrt{\frac{\gamma}{2}} \left(\langle\hat{\sigma}_1^-\rangle + i\langle\hat{\sigma}_2^-\rangle \right), \\
        &\langle\hat{a}_\textrm{R}\rangle = \langle \hat a_\mathrm{R}^\mathrm{in}  \rangle +  \sqrt{\frac{\gamma}{2}} \left(\langle\hat{\sigma}_1^-\rangle -i\langle\hat{\sigma}_2^-\rangle\right).
\end{split}
\end{equation}

We use numerical master equation simulations to determine the transmission amplitude $S_{21} = \langle\hat a_\mathrm{R}\rangle/\langle\hat a_\mathrm{R}^\textrm{in}\rangle$ as a function of probe detuning $\delta$, emitter detuning $\Delta$, and the total emitter-emitter coupling $J_\Sigma$ as shown in Fig.~\ref{fig:fig2sim}.
We also simulate the transmission of a resonant probe as a function of probe power as shown in Fig.~\ref{fig:fig2}c.
The simulations mirror our calibration spectroscopic experiments, indicating that this model captures the steady-state dynamics of the driven two-qubit system.

We experimentally find the $|J_\Sigma|=0$ point by varying the frequency of the tunable coupler $C_{12}$, which in turn adjusts the net interaction between the two emitter qubits.
In particular, we measure $|S_{21}|$ for when the input probe is resonant with both emitter qubits $\delta = 0$ while sweeping the frequency of $C_{12}$, as shown in Fig.~\ref{fig:cancel}a.
The frequency for which $|S_{21}|$ is maximized corresponds to the operating point when $|J_\Sigma|$ is minimized, and is ideally zero.
To see this more clearly, we map the frequency of $C_{12}$ onto $J_\Sigma$ (see Refs.~\cite{yan2018,sung2020}) in Fig.~\ref{fig:cancel}b.
For the plotted range ($J_\Sigma \in \left[-\gamma/2,\gamma/2\right]$, we can clearly see that $|S_{21}|$ is maximized when $J_\Sigma = 0$.
Note that the $|S_{21}|$ slightly exceed unity at its maximum value, which we attribute to impedance mismatches in our experimental setup \cite{Khalil2012,Probst2015}.
Finally, we plot $|S_{21}|$ as a function of the probe detuning $\Delta$ from the emitter qubits in Fig.~\ref{fig:cancel}c.
We show four representative traces for different values of $|J_\Sigma|$. When $|J_\Sigma|>\gamma/2$, we observe a splitting from the hybridization of the qubits.
However, when $|J_\Sigma|<\gamma/2$ we simply observe a single dip that shallows as the coupling decreases.

Experimentally, the near-unity transmission measured in Fig. 2 is only possible when the separation between qubits is close to $\lambda/4$. We note that both the transmission measurements and the directional emission protocol are quite insensitive to small deviations from $\Delta x = \lambda/4$, as shown in \cite{Gheeraert2020}, which is a feature of the protocol.

We also note that the directional emission fidelity is much more sensitive to the dephasing rates of the qubits, because the directionality is determined by the interference during the two-qubit emission event. Therefore, operating at a frequency with lower sensitivity to flux-noise can result in higher fidelities, even if that frequency deviates slightly from the $\lambda/4$ condition. In our case, we have designed the frequency of the flux-noise sweet spots of the qubits to be as close to the $\lambda/4$ condition as possible.

\begin{center}
    \textbf{Parametric Exchange Interactions}
\end{center}

\begin{figure*}[t!]
    \centering
    \includegraphics[width=7in]{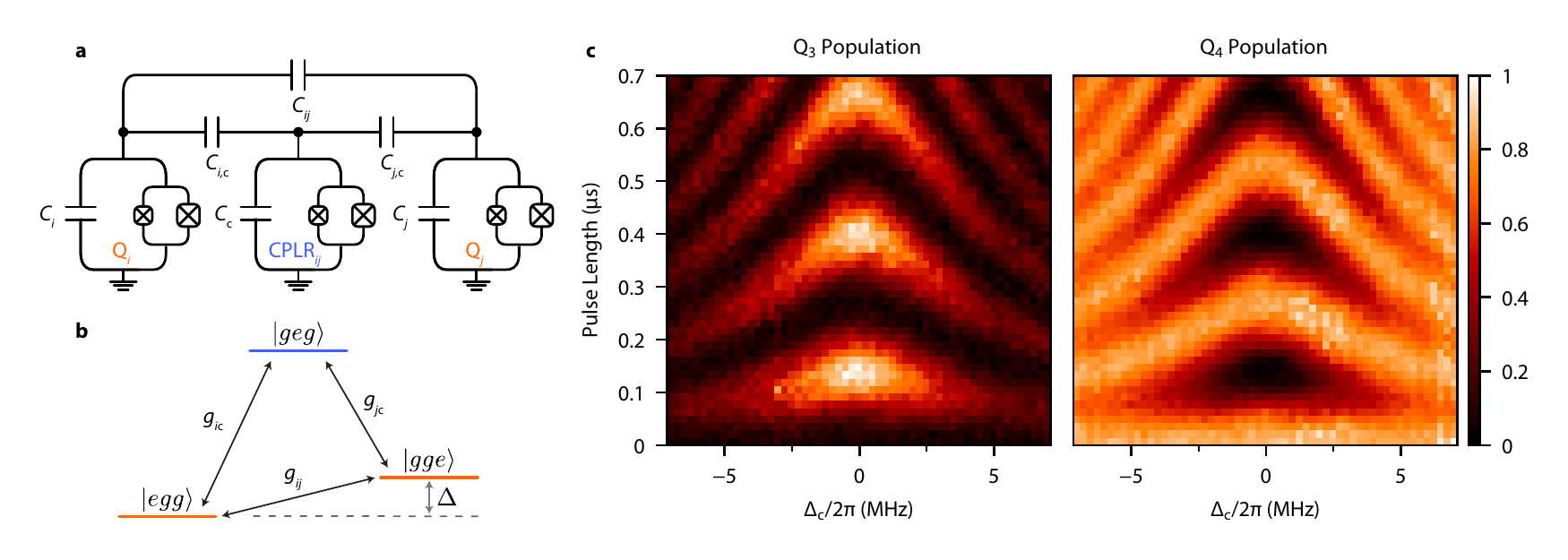}
    
    \caption{\textbf{Parametric interactions with a tunable coupler. } a) The circuit diagram of a system with a tunable coupler CPLR$_{ij}$ capacitively coupled two tunable transmon qubits $\textrm{Q}_i$ and $\textrm{Q}_j$. The coupling capacitance between the qubits is $C_{ij}$ and the coupling capacitances between each qubit and the coupler is $C_{i,c}$ ($C_{j,c}$). b) The single-excitation manifold level diagram of the system. The coupler frequency $\omega_c$ is far-detuned from the frequencies of the qubits $\omega_i$ and $\omega_j$, and the two qubits are slightly detuned from each other by $\Delta = \omega_j - \omega_i$. The capacitance between the qubits $C_{ij}$ mediates a direct coupling with strength $g_{ij}$. The capacitances between the coupler and each qubit give rise to couplings between each qubit-coupler pair at the rates $g_{ic}$ and $g_{jc}$. c) The measured population exchange between qubits $\textrm{Q}_3$ and $\textrm{Q}_4$ as a function of the parametric modulation pulse length and frequency offset $\Delta_c = \delta_c - \Delta$, where $\delta_c$ is the frequency of the modulation.}
    \label{fig:chevron}
\end{figure*}

The exchange interactions used in the main text were mediated by the parametric modulation of the tunable coupler frequencies. To see this, consider two qubits at frequencies $\omega_i$ and $\omega_j$, and a tunable coupler at frequency $\omega_c$, as illustrated in Fig.~\ref{fig:chevron}a. Each qubit is capacitively coupled to the tunable coupler at rates $g_{ic}$ and $g_{jc}$, and to each other at rate $g_{ij}$. The coupler is far-detuned from both qubit frequencies $\omega_c - \omega_{i/j} \gg g_{ic},g_{jc}, g_{ij}$, and the qubits are slightly detuned by $\Delta  = \omega_j - \omega_i$. The Hamiltonian of the system in the rotating frame of the qubit Q$_i$ is~\cite{yan2018}
\begin{equation}
    \hat{H} = \Delta\sigma_j^+ \sigma_j^- + \left(\frac{g_{ic}g_{jc}}{\delta} + g_{ij}\right)(\sigma_i^+ \sigma_j^- + \sigma_i^- \sigma_j^+),
\end{equation}
where $\delta = 2\left( \frac{1}{\omega_i - \omega_c} + \frac{1}{\omega_j- \omega_c}\right)^{-1} = 2\left(\frac{1}{\delta_i} + \frac{1}{\delta_j}\right)^{-1}$. The coupling rates $g_{ic}, g_{jc},$ and $g_{ij}$ are determined by the the qubit frequencies and the capacitances $C_{ij}$, $C_c$, $C_i/C_j$, and $C_{j,c}/C_{i,c}$ as defined in Fig.~\ref{fig:chevron}a:
\begin{equation}
\begin{split}
    &g_{ij} \approx \frac{1}{2}\left( \frac{C_{ij}}{\sqrt{C_iC_j}} + \frac{C_{i,c}C_{j,c}}{\sqrt{C_iC_jC_c^2}}  \right)\sqrt{\omega_i\omega_j} = \frac{\tilde{C}_{ij}}{2}\sqrt{\omega_i\omega_j},\\
    &g_{nc} \approx \frac{C_{n,c}}{2\sqrt{C_nC_c}}\sqrt{\omega_n\omega_c}= \frac{\tilde{C}}{2}\sqrt{\omega_n\omega_c} \indent \indent n = i,j.
\end{split}
\end{equation}
Here, we assume that Q$_i$ and Q$_j$ are identical qubits, with equal self-capacitances $C_i = C_j$ and capacitances to the coupler $C_{i,c} = C_{j,c}$. Substituting these expressions for the coupling rates into the Hamiltonian gives
\begin{equation}
    \hat{H} = \Delta\sigma_j^+ \sigma_j^- + \sqrt{\omega_i\omega_j}\left(\frac{\tilde{C}^2\omega_c}{4\delta} + \frac{\tilde{C}_{ij}}{2} \right)
    (\sigma_i^+ \sigma_j^- + \sigma_i^- \sigma_j^+).
\end{equation}

Next, we modulate the frequency of the tunable coupler $\omega_c = \omega_{c0} + A\cos{\Delta t}$. In practice, this is realized by modulating the flux applied into the SQUID loop of the coupler. Since $\omega_c\gg \omega_i, \omega_j$, we can approximate the  total detuning as $\delta \approx \delta_i = \omega_i - \omega_c$. Assuming the amplitude of the coupler frequency modulation $A \ll \delta_i$, we separate the qubit coupling into a static component and a time-varying component,
\begin{equation}
    \hat{H} = \Delta\sigma_j^+ \sigma_j^- + \sqrt{\omega_i\omega_j}\left(\frac{\tilde{C}^2\omega_{c0}}{4\delta} + \frac{\tilde{C}_{ij}}{2} + \frac{\tilde{C}^2 A\cos{\Delta t}}{4\delta}  \right)
    (\sigma_i^+ \sigma_j^- + \sigma_i^- \sigma_j^+).
\end{equation}
Finally, we rotate into the frame of the qubit detuning $\Delta$ and neglect the fast rotating terms. This approximation holds as long as the effective coupling rate $g_\mathrm{eff} \ll \Delta$. The final time-independent Hamiltonian is given by
\begin{equation}
    \hat{H} = \frac{A\tilde{C}^2 \sqrt{\omega_i\omega_j}}{8\delta}
    (\sigma_i^+ \sigma_j^- + \sigma_i^- \sigma_j^+).
\end{equation}
This Hamiltonian shows that the two detuned qubits $\textrm{Q}_i$ and $\textrm{Q}_j$ are effectively coupled at rate $g_\mathrm{eff} = A\tilde{C}^2 \sqrt{\omega_i\omega_j}/8\delta $. We show in Fig.~\ref{fig:chevron}c measurements of the chevron pattern for population exchange between qubits $\textrm{Q}_3$ and $\textrm{Q}_4$ mediated by a parametric exchange interaction. Note that we can also vary we vary the effective coupling rate as a function of time by varying the frequency modulation amplitde $A(t)$. This feature can be used to shape the wavepacket of the emitted photon, which will be necessary in future work for perfect absorption of the emitted photons~\cite{Kurpiers2018,Gheeraert2020,Forn-Diaz2017,reuer2021realization}.

\begin{center}
    \textbf{Time-Domain Measurements}
\end{center}

In Fig.~\ref{fig:fig3}b/c of the main text, we showed the temporal dynamics of the photon wavepacket. Here, we analytically derive the shape of the wavepacket. We prepared the data qubits in the state $|\psi_\mathrm{qb}^\pm\rangle = (|gg\rangle + |\psi^\pm\rangle)/\sqrt{2}$. The state of the data qubits was then transferred to the emitter qubits with parametric exchange interactions as part of the photon release protocol. The four-qubit master equation that describes this system is written as

\begin{equation}
    \partial_t \hat{\rho} = -i[\hat H, \hat \rho] + \sum_{k = 1,2} \hat c_k \hat \rho \hat c_k^{\dagger} -\frac{1}{2}\{\hat c_k^\dagger \hat c_k,\hat \rho\},
    \label{eq: master equation}
\end{equation}

\begin{figure*}[t!]
    \centering
    \includegraphics[width=7in]{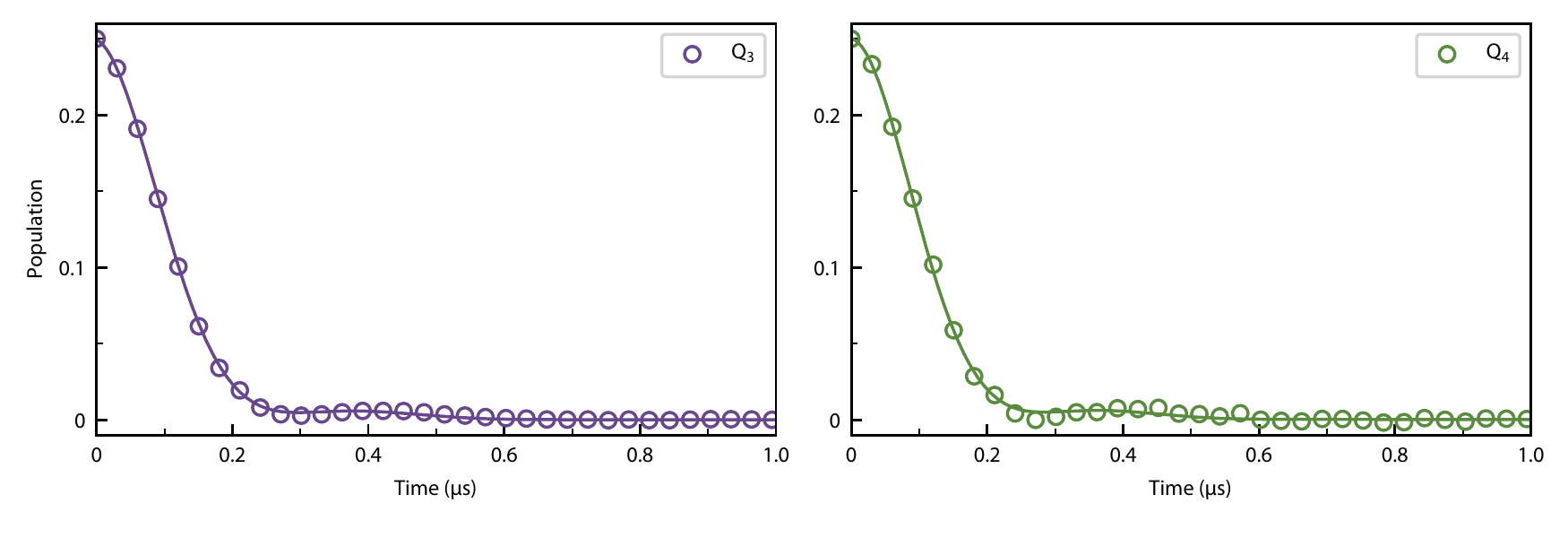}
    
    \caption{\textbf{Excited state population of data qubits during photon emission.} We utilize dispersive readout to measure the population of the data qubits during the photon release protocol immediately after the initialization of the data qubits into the state $|\psi_\mathrm{qb}^{\pm}\rangle = |gg\rangle/\sqrt{2} + (|eg\rangle + e^{\pm i\frac{\pi}{2}}|ge\rangle)/2$. The theoretical fit of the population is presented as a solid line in both plots.}
    \label{fig:dataqtime}
\end{figure*}
where 
\begin{equation}
    \hat H = g_\mathrm{eff}\,(\hat \sigma^-_3 \hat \sigma^+_1+ \hat \sigma^+_3 \hat \sigma^-_1) + g_\mathrm{eff}\,(\hat \sigma^-_4 \hat \sigma^+_2+ \hat \sigma^+_4 \hat \sigma^-_2),
\end{equation}
is the system's Hamiltonian and 
\begin{equation}
    \hat c_k\in\{\sqrt{\gamma} \hat \sigma^-_1, \sqrt{\gamma}\hat \sigma^-_2\},
\end{equation}
are the collapse operators. The raising and lowering operators of each qubit Q$_i$ is denoted as $\hat \sigma_i^{\pm}$, where $i \in \{1,2,3,4\}$ is the qubit number as defined in the main text.

We work in the superoperator representation of Eq. \ref{eq: master equation}. Vectorizing the density matrix as~$\hat\rho\to|\rho\rangle\rangle$, we rewrite the master equation as 
\begin{equation}
    \partial_t|\rho\rangle\rangle = \mathcal{\hat L}|\rho\rangle\rangle,
    \label{eq: superop master equation}
\end{equation}
where~$\mathcal{\hat L}$ is the Liouvillian superoperator~\cite{havel2003robust}
\begin{equation}
\hat{\mathcal{L}}=-i\left( \mathds{1} \otimes \hat{H}-\hat{H}^{T} \otimes  \mathds{1} \right)+\sum_{k} \hat{c}_{k}^{*} \otimes \hat{c}_{k}-\frac{1}{2}\left( \mathds{1}  \otimes \hat{c}_{k}^{\dagger} \hat{c}_{k}+\hat{c}_{k}^{T} \hat{c}_{k}^{*} \otimes  \mathds{1}\right)
    \label{eq: L}
\end{equation}
We can formally express the solution to Eq. \ref{eq: superop master equation} as 
\begin{equation}
    |\rho(t)\rangle\rangle = \hat S(t)|\rho(0)\rangle\rangle,
    \label{eq: superop master equation solution}
\end{equation}
where
$
    \hat S(t) = \exp(\mathcal{\hat L}\,t)
    \label{eq: S}
$
is the quantum channel described by the original master equation in superoperator form. In this simplified model, the subspace formed by Q$_1$ and Q$_3$ is not coupled to the subspace formed by Q$_2$ and Q$_4$, which allows us to write the Liouvillian superoperator as $\mathcal{\hat L} = \mathcal{\hat L}_{13} +  \mathcal{\hat L}_{24}$, where $\mathcal{\hat L}_{13(24)}$ is the Liouvillian superoperator of each two-qubit subsystem. Because $\mathcal{\hat L}_{13}$ and $\mathcal{\hat L}_{24}$ commute, we can factorize the quantum channel as
\begin{equation}
    \hat S(t) = \exp(\mathcal{\hat L}_{13} \,t)\cdot\exp(\mathcal{\hat L}_{24}\,t).
    \label{eq: S factorized}
\end{equation}
We solve for the density matrix of the four-qubit system $\hat \rho$, which we use to compute system observables, i.e.~$\langle \hat{O}\rangle=\mathrm{Tr}[\hat \rho\hat{O}]$. First, we examine the data qubit population as a function of time during the photon release. We obtain the analytical expression for the excited state population of each data qubit as a function of time:
\begin{equation}
        \rho_{33}^{(d)}(t) = \rho_{44}^{(d)}(t) =  \frac{e^{-\frac{\gamma}{2}t}}{16\Gamma^2} \left[ (\gamma^2 - 8g_\mathrm{eff}^2) \cosh{\left(\Gamma t\right)} + 2\gamma\Gamma \sinh{\left(\Gamma t\right)} -8g_\mathrm{eff}^2 \right],\\
        \label{eq:population}
\end{equation}
where we define $\Gamma = 2\sqrt{\left(\frac{\gamma}{4}\right)^2 -g_\mathrm{eff}^2}$. We use Eq.~\ref{eq:population} to fit the dispersive readout measurement of the data qubits during the photon release protocol, shown in Fig. \ref{fig:dataqtime}. The decay in population here corresponds to its transfer into the emitter qubits and subsequent release into the waveguide.

To obtain the temporal wavepacket of the emission field amplitude, we use the input-output relations:
\begin{equation}
\begin{split}
        &\langle\hat{a}_\textrm{L}\rangle = \sqrt{\frac{\gamma}{2}} \left(\langle\hat{\sigma}_1^-\rangle + \langle\hat{\sigma}_2^-\rangle e^{i\frac{\pi}{2}}\right), \\
        &\langle\hat{a}_\textrm{R}\rangle = \sqrt{\frac{\gamma}{2}} \left(\langle\hat{\sigma}_1^-\rangle + \langle\hat{\sigma}_2^-\rangle e^{-i\frac{\pi}{2}}\right).
\end{split}
\end{equation}
Here, we assume that there is no input into the waveguide $\langle\hat a_\mathrm{L/R}^\mathrm{in}\rangle = 0$. Using the solution to the master equation, we compute the wavepacket shape, given the initial state of the emitter qubits:
\begin{equation}
    \begin{split}
    &|\psi_\mathrm{qb}^-\rangle = \frac{|gg\rangle + |\psi^-\rangle}{\sqrt{2}} \indent \to \indent \langle \hat{a}_\textrm{L}\rangle = -\frac{g_\mathrm{eff}\sqrt{\gamma}}{\Gamma}e^{-\frac{\gamma}{4}t}\sinh{\left(\frac{\Gamma}{2}t\right)}, \indent \indent \langle\hat a_\mathrm{R}\rangle = 0,\\
    &|\psi_\mathrm{qb}^+\rangle = \frac{|gg\rangle + |\psi^+\rangle}{\sqrt{2}} \indent \to \indent \langle \hat{a}_\textrm{R}\rangle = \frac{g_\mathrm{eff}\sqrt{\gamma}}{\Gamma}e^{-\frac{\gamma}{4}t}\sinh{\left(\frac{\Gamma}{2}t\right)},~~ \indent \indent \langle\hat a_\mathrm{L}\rangle = 0,
    \end{split}
\end{equation}
which is used to fit the photon field amplitudes in Fig. \ref{fig:fig3}b/c and extract the effective coupling between each emitter/data qubit pair $g_\mathrm{eff}/2\pi \approx 1.28$ MHz.


\end{document}